\documentclass[apjl]{emulateapj}
\usepackage{amssymb}
\usepackage{times}
\usepackage{epsfig}
\usepackage{graphics}

\usepackage{amsmath}

\def\swift{{\it Swift}}
\def\nustar{{\it NuSTAR}}
\def\fermi{{\it Fermi}}

\begin{document}
NuSTAR observations of GRB 1304127A establish 
\title{\nustar\ Observations of GRB\,$130427$A establish a single component synchrotron afterglow origin for the late optical to multi-GeV emission}

\author{C.~Kouveliotou\altaffilmark{1,*},
J.~Granot\altaffilmark{2,*}, J.~L.~Racusin\altaffilmark{3,*},
E.~Bellm\altaffilmark{4}, G.~Vianello\altaffilmark{5},
S.~Oates\altaffilmark{6}, C.~L.~Fryer\altaffilmark{7},
S.~E.~Boggs\altaffilmark{8}, F.~E.~Christensen\altaffilmark{9},
W.~W.~Craig\altaffilmark{8,10}, C.~D.~Dermer\altaffilmark{11},
N.~Gehrels\altaffilmark{3}, C.~J.~Hailey\altaffilmark{12},
F.~A.~Harrison\altaffilmark{4}, A.~Melandri\altaffilmark{13},
J.~E.~McEnery\altaffilmark{3}, C.~G.~Mundell\altaffilmark{14},
D.~K.~Stern\altaffilmark{15}, G. Tagliaferri\altaffilmark{13},
W.~W.~Zhang\altaffilmark{3} }

\altaffiltext{1}{Astrophysics Office/ZP12, NASA Marshall Space Flight Center, Huntsville, AL35812, USA}
\altaffiltext{2}{Department of Natural Sciences, The Open University of Israel, 1 University Road, P.O. Box 808, Ra'anana 43537, Israel}
\altaffiltext{3}{NASA Goddard Space Flight Center, Greenbelt, MD 20771, USA}
\altaffiltext{4}{Cahill Center for Astronomy and Astrophysics, California Institute of Technology, 1200 E.California Blvd., Pasadena, CA 91125, USA}
\altaffiltext{5}{W. Hansen Experimental Physics Laboratory, Kavli Institute for Particle Astrophysics andCosmology, Department of Physics and SLAC National Accelerator Laboratory, StanfordUniversity, Stanford, CA 94305, USA}
\altaffiltext{6}{Mullard Space Science Laboratory, University College London, Holmbury St. Mary, Dorking,Surrey RH5 6NT, UK}
\altaffiltext{7}{CCS-2, Los Alamos National Laboratory, Los Alamos, NM 87545, USA}
\altaffiltext{8}{Space Sciences Laboratory, University of California, Berkeley, CA 94720, USA}
\altaffiltext{9}{DTU Space-National Space Institute, Technical University of Denmark, Elektrovej 327, 2800Lyngby, Denmark}
\altaffiltext{10}{Lawrence Livermore National Laboratory, Livermore, CA, 94550, USA}
\altaffiltext{11}{Code 7653, National Research Laboratory, Washington, DC 20375-5352, USA}
\altaffiltext{12}{Columbia Astrophysics Laboratory, Columbia University, New York, NY 10027, USA}
\altaffiltext{13}{INAF-Osservatorio Astronomico di Brera, via E. Bianchi 46, I-23807 Merate, Italy}
\altaffiltext{14}{Astrophysics Research Institute, Liverpool John Moores University, 146 Brownlow Hill,Liverpool Science Park, Liverpool L3 5RF, UK}
\altaffiltext{15}{Jet Propulsion Laboratory, California Institute of Technology, Pasadena, CA 91109, USA}
\altaffiltext{*}{Contact author}

\email{chryssa.kouveliotou@nasa.gov}

\begin{abstract}
GRB\,130427A occurred in a relatively nearby galaxy; its prompt
emission had the largest GRB fluence ever recorded. The afterglow of
GRB\,130427A was bright enough for the Nuclear Spectroscopic Telescope
ARray ({\it NuSTAR}) to observe it in the 3$\,$--$\,79\;$keV energy
range long after its prompt emission ($\sim$1.5 and 5$\;$days). This
range, where afterglow observations were previously not possible,
bridges an important spectral gap. Combined with {\it Swift, Fermi}
and ground-based optical data, {\it NuSTAR} observations unambiguously
establish a single afterglow spectral component from optical to
multi-GeV energies a day after the event, which is almost certainly
synchrotron radiation. Such an origin of the late-time \fermi/LAT
$>$10$\;$GeV photons requires revisions in our understanding of
collisionless relativistic shock physics.
\end{abstract}


\section{Introduction}

Gamma-Ray Bursts (GRBs) release within seconds to minutes more
high-energy photons than any other transient phenomenon
\citep{book}. Their prompt gamma-ray emission is followed by a
long-lived (typically weeks to months) afterglow, visible from radio
to X-rays.  The afterglow emission is attributed to synchrotron
radiation from relativistic electrons accelerated in the shock
produced as the explosion plows into the circumstellar medium. The
afterglow synchrotron origin is supported by their broadband spectra
\citep{GS02,Galama98} and polarization measurements \citep{Covino04}.

GRB$\;$130427A triggered the {\it Fermi}/Gamma-ray Burst Monitor (GBM)
at 07:47:06.42 UT on 2013 April 27 \citep{kienlin13}. The intensity
and hardness of the event fulfilled the criteria for an autonomous
slew maneuver to place the burst within the {\it Fermi}/Large Area
Telescope (LAT) field-of-view. Its exceedingly bright prompt emission
was also detected by other satellites ({\it AGILE}:
\citealt{Verrecchia13}, {\it Konus-Wind}: \citealt{Golenetskii13},
{\it RHESSI}: \citealt{Smith13}, {\it Swift}: \citealt{Maselli13}) and
enabled multiple ground- and space-based follow-up observations,
allowing for rapid accurate determination of the event location and
distance at redshift $z=0.340$
\citep{Levan13}, as well as extensive broad band afterglow monitoring
from radio to $\gamma$-rays. The extreme X-ray and $\gamma$-ray
energetics of the burst are described in detail in
\citet{Preece13,Ackermann13,Maselli13Sc}. The record-breaking duration
of the LAT afterglow ($\sim$0.1--100$\;$GeV), which lasted
almost a day after the GBM trigger, placed GRB$\;$130427A at the top of
the LAT GRBs in fluence \citep{Ackermann13}.

The extreme intensity, accurate distance measurement and relative
closeness of GRB$\;$130427A, made it an ideal candidate for follow-up
observations with \nustar$\;$\citep{Harrison13}. Here we describe our
\nustar\ afterglow observations taken during two epochs
(\S$\;$\ref{sec:nustar}), combined with data from \fermi/LAT, \swift, and
optical observatories. We describe in \S$\;$\ref{sec:fermi} the
derivation of the \fermi/LAT extrapolation and upper limits during the
\nustar\ epochs. In \S$\;$\ref{sec:broadband} we present afterglow multi-wavelength
fits, and discuss our results in \S$\;$\ref{sec:disc}.

\section{\nustar\ Observations}
\label{sec:nustar}

\nustar\ was launched on
2012 June 13; the instrument's two telescopes utilize a new generation
of hard X-ray optics and detectors to focus X-rays in the range
3$\;$--$\;$79$\;$keV. We observed GRB$\;$130427A at three epochs,
starting approximately 1.2, 4.8 and 5.4 days after the GBM trigger,
for 30.5,$\;$21.2, and 12.3$\;$ks (live times).  We detected the
source in all epochs, obtaining for the first time X-ray observations
of a GRB afterglow above 10$\;$keV. The
\nustar\ data thus provide an important missing spectral link between
the \swift/X-Ray Telescope (XRT) observations (0.3$\;$--$\;$10$\;$keV)
\citep{Maselli13Sc} and the \fermi/LAT observations ($>100\;$MeV)
\citep{Ackermann13}.

We processed the data with {\it HEASOFT} 6.13 and the \nustar\ Data
Analysis Software (NuSTARDAS) v.~1.1.1 using CALDB version
20130509. We extracted source lightcurves and spectra from circular
regions with 75$^{\prime\prime}$ radius from both \nustar\ modules for
the first epoch and 50$^{\prime\prime}$ radius for the second and
third epochs.  We used circular background regions (of
150$^{\prime\prime}$, 100$^{\prime\prime}$, and 100$^{\prime\prime}$
radius for each epoch, respectively) located on the same \nustar\
detector as the GRB. Hereafter, we combine the second and third
\nustar\ epochs, which were very close in time, to increase the
signal-to-noise ratio, and refer to it as the second epoch.

\begin{figure}
\begin{center}
\includegraphics[angle=90,width=1.0\columnwidth]{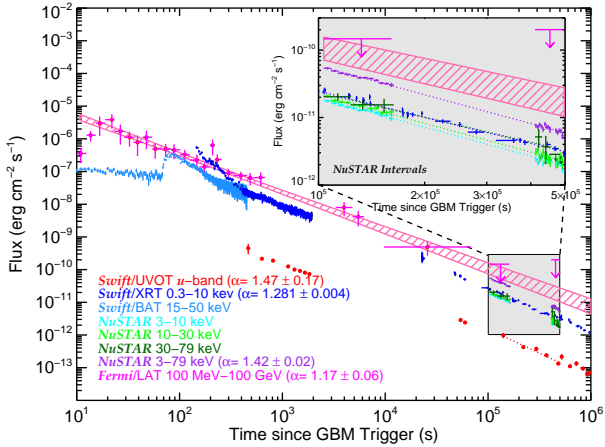}
\caption{Lightcurves of GRB$\;$130427A. \citep{Maselli13Sc,Ackermann13}.  
All errors are 1$\sigma$. The inset zooms in on the \nustar\
epochs. The LAT ULs are shown as arrows and the LAT extrapolated
region as a shaded rectangle (1$\sigma$). Numbers in parentheses are
indices of power-law fits during the \nustar\ epochs.}
\label{fig:lc}
\end{center}
\end{figure}

Fig.$\;$\ref{fig:lc} demonstrates the temporal behavior of the
multi-wavelength afterglow flux of GRB$\;$130427A. Here we have included
data from \swift/XRT, \swift/UVOT, and \fermi/LAT. We also include the
extrapolated \fermi/LAT lightcurve derived as described in
\S$\;$\ref{sec:fermi}. The weighted average of the decay rates during the
two \nustar\ epochs (single power-law fits) is $\alpha=1.3$ from
optical to GeV (see also the figure inset, and the indices next to
each instrument in Fig.$\;$\ref{fig:lc}). We discuss the implications of the
temporal results in \S$\;$\ref{sec:disc}.

\section{\fermi\ Observations}
\label{sec:fermi}

The \fermi/LAT detected GRB$\;$130427A up to almost a day after the
trigger time (Fig.$\;$\ref{fig:lc}; \citealt{Ackermann13}). \fermi/LAT
was also observing during both \nustar\ epochs but did not detect the
source.  We analyzed the ``Pass$\;$7'' data with the Fermi Science
Tools v9r31p1 and the P7SOURCE\_V6 version of the instrument response
functions, and using the public Galactic diffuse model and the
isotropic spectral template available at
http://fermi.gsfc.nasa.gov/ssc/.  For each epoch, we selected all the
events within a Region Of Interest (ROI) with a radius of 10$^{\circ}$
around the position of the GRB, excluding times when any part of the
ROI was at a zenith angle $>100{^\circ}$. The latter requirement
greatly reduces contamination from the diffuse gamma-ray emission
originating from the Earth's upper atmosphere, peaking at a zenith
angle of $\sim$110$^\circ$.

\subsection{\fermi/LAT spectra and upper limits} 

For each epoch we performed an unbinned likelihood analysis over the
whole energy range (0.1$\;$--$\;$100$\;$GeV), using a model composed
of the two background components (Galactic and isotropic) and a point
source with a power-law (PL) spectrum (the GRB), plus the contribution
from all the known gamma-ray point sources in the ROI
\citep{Nolan12}. We did not obtain a detection in either epoch, and so
we computed upper limits (ULs). We froze the normalization of the
background components, and fixed the photon index of the GRB model to
$2.17$, which is the best fit value from the smoothly-broken power-law
(SBPL) fit during the first \nustar\ epoch as reported in
\S$\;$\ref{sec:broadband} (the ULs change by less than 10\% for any choice
of the photon index between $2$ and $2.5$). We then independently fit
the GRB model in 3 energy bands ($0.1-1$, $1-10$ and
$10-100\;$GeV), using an unbinned profile likelihood method to derive
the corresponding 95\% LAT ULs \citep{Ackermann12}.  The information
contained in such ULs is important to constrain the spectrum, but
cannot be handled by a standard fitting procedure. We, therefore, turn
to an alternative (but equivalent) method to include the LAT
observations in a broadband spectral fit.  We obtained the count
spectrum of the observed LAT signal (source+background) using
\textit{gtbin}, and the background spectrum using \textit{gtbkg},
which computes the predicted counts from all the components of the
best fit likelihood model except the GRB. Since there is no
significant excess above the background, the two spectra are
compatible within the errors, although they are not identical. We also
ran \textit{gtrspgen} to compute the response of the instrument in the
interval of interest, and loaded these files in XSPEC$\;$v.12.7. This
software compares the observed net counts to the number of counts
predicted by the model folded with the response of the instrument. By
minimizing a statistic based on the Poisson probability we can treat
equivalently a spectrum containing a significant signal, and a
spectrum which is compatible with being just background. While the
former will constrain the model to pass through the data points, the
latter will constrain it to predict a number of counts above
background compatible with zero. The best-fit model obtained using the
LAT spectra computed in this way is, as expected, below the ULs
computed with the profile likelihood method.

\subsection{Extrapolation of the \fermi/LAT lightcurve}

The high-energy ($>100\;$MeV) photon and energy flux lightcurves are
well described by a broken power law (BPL) and PL, respectively, as
reported in \citet{Ackermann13}. To extrapolate such lightcurves to
the \nustar\ epochs we adopted a general approach, based on the
well-known Markov Chain Monte Carlo technique, which takes into
account the uncertainties on the best fit parameters along with all
their correlations, as follows.

\begin{figure}
\begin{center}
\includegraphics[width=0.9\columnwidth]{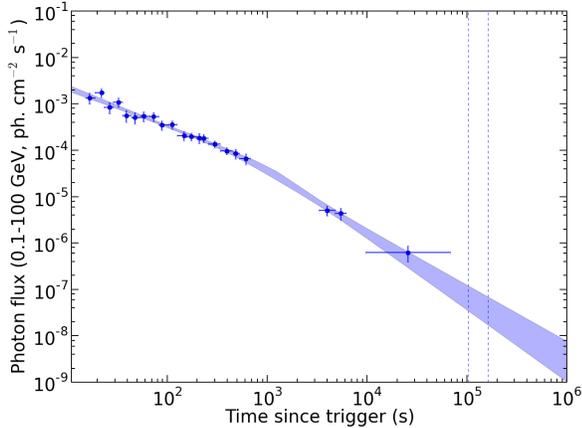}
\caption{The decaying part of the \fermi/LAT photon flux lightcurve 
of the afterglow of GRB$\;$130427A (100$\;$MeV$\;$--$\;$100$\;$GeV;
\citealt{Ackermann13}). The shaded blue region marks the 1$\sigma$
contour for the best-fit BPL model, while the dashed lines indicate
the start and stop time for the first \nustar\ epoch.  Only data
points used for the BPL fit (i.e., after $T_{0}+16\;$s) are included.
}
\label{fig:lat}
\end{center}
\end{figure}

Each data point in Fig.$\;$\ref{fig:lat} represents a photon flux
derived from a likelihood fit with 1$\sigma$ confidence intervals
\citep{Ackermann13}. Hence, we can assume a Gaussian joint likelihood $L$ and
minimize the corresponding $-\log(L)$ to find the best-fit parameters,
which is equivalent to a standard least-squares fit (or to minimize
$\chi^2$). We can then apply the Bayes rule that the posterior
distribution for the parameters is directly proportional to the prior
distribution multiplied by the likelihood. If we take an uninformative
prior, then the posterior distribution is directly proportional to the
likelihood itself. Therefore, sampling the likelihood function with a
Markov Chain Monte Carlo technique is equivalent to sampling the
posterior distribution. By using {\it e.g.,} the \citet{GW10}
algorithm, we can then obtain many sets of parameters distributed as
in the posterior distribution, with all the relations between them
taken into account. Using these sets of parameters, $p_{\rm i}$, we can
build a distribution of a certain quantity of interest
$f(p_{\rm i})$. Taking the median and the relevant percentiles of the
distribution we can then extract a measure of $f$ and its 1$\sigma$
confidence interval. In this way, we computed the shaded region in
Fig.$\;$\ref{fig:lat} and the expected flux only in the first
\nustar\ epoch, which starts shortly after the last detection from 
\fermi/LAT. The second \nustar\ epoch started too late for any 
extrapolation to be meaningful.

Fig.$\;$\ref{fig:lat} exhibits the \fermi/LAT photon flux lightcurve
with 1$\sigma$ confidence intervals derived with such method. We used
the same method to compute the flux extrapolation for the first
\nustar\ epoch (the magenta dashed cross in Fig.$\;$\ref{fig:sed}).

\section{Broadband Afterglow}
\label{sec:broadband}

We extracted lightcurves and spectra during the \nustar\ epochs from
\swift/Ultra-Violet/Optical Telescope (UVOT), and \swift/XRT using the
standard HEASOFT reduction pipelines and the \swift/XRT team
repository \citep{Evans09}, as well as Liverpool Telescope data using
in-house software \citep{Maselli13Sc}. For the first epoch, we compare
the extrapolation of the LAT temporal and spectral behavior
\citep{Ackermann13} to our multi-wavelength lightcurves and spectra.

\begin{figure}
\begin{center}
\includegraphics[width=1.0\columnwidth]{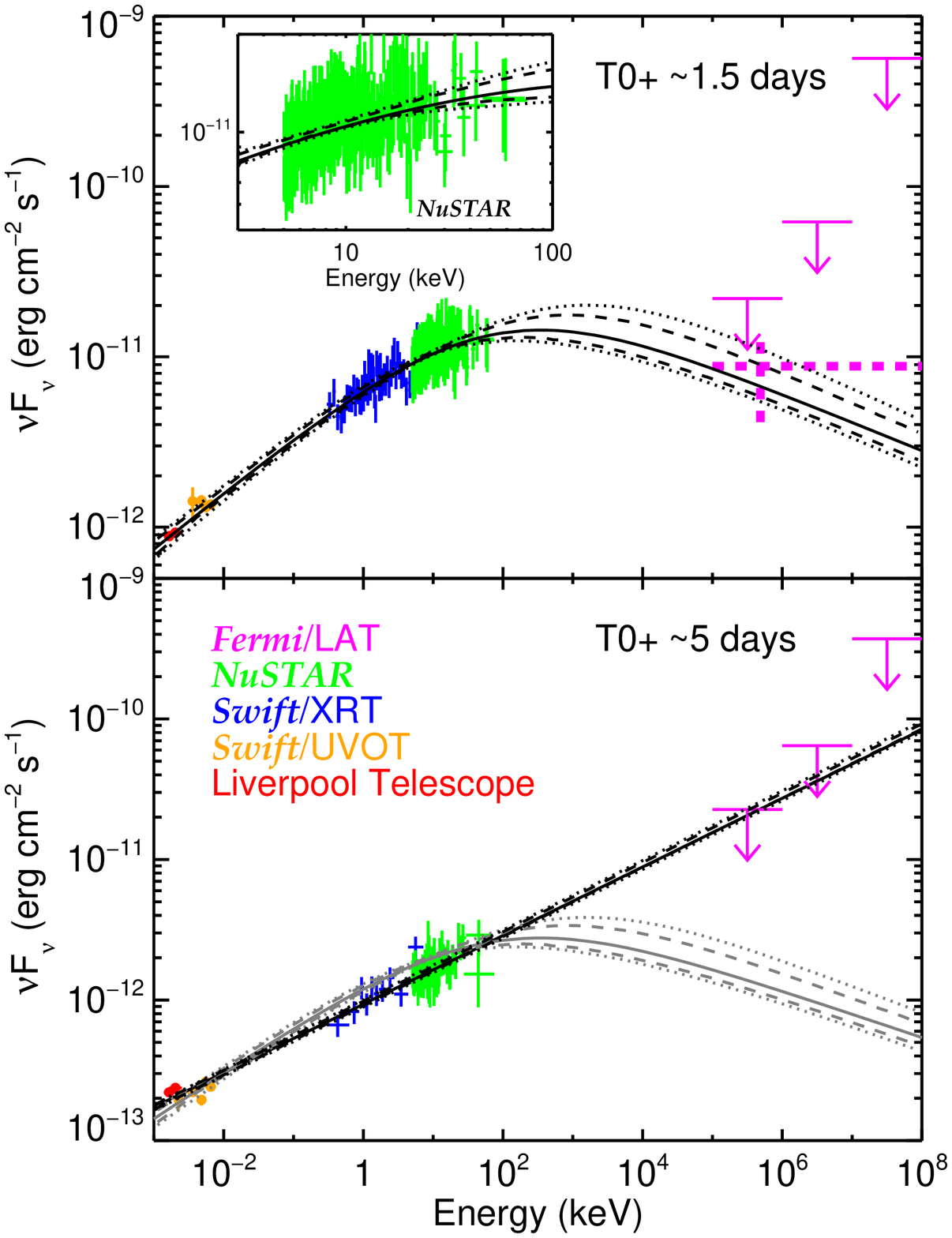}
\caption{The optical to GeV spectrum of GRB$\;$130427A 
fit with a SBPL synchrotron model \citep{GS02}. Broad-band SEDs are
shown during the first ({\it top-panel}) and the second ({\it
bottom-panel}) \nustar\ epochs. The \fermi/LAT ULs are shown as arrows
and the extrapolation of the LAT flux light curve is shown as a dashed
magenta cross (only during the first epoch). The second epoch ({\it
bottom-panel}) is fit with a PL ({\it black lines}); the fit to the
first epoch is scaled down and superposed on the second epoch data for
comparison ({\it in gray}).The optical/UV/XRT data are corrected for
absorption and Galactic extinction. All data point errors are
1$\sigma$; the LAT ULs are 2$\sigma$; the error contours are 2$\sigma$
({\it dashed lines}) and 3$\sigma$ ({\it dotted lines}).}
\label{fig:sed}
\end{center}
\end{figure}

Fig.$\;$\ref{fig:sed} shows two Spectral Energy Distributions (SEDs)
spanning from optical (i' band) to $\gamma$-rays ($\sim$GeV). We first
fit both epochs independently (excluding \fermi/LAT data) with two
functional forms (Table~1) -- single PL and BPL -- each multiplied by
models for both fixed Galactic and free intrinsic (host) extinction
(zdust)\footnote{The Small Magellanic Cloud (SMC) extinction curve
fits our data best and we use it exclusively for continuity.} and
absorption (phabs), respectively, and a free cross-calibration
constant. We find that both epochs can be fit with a PL; however, the
second epoch fit is better ($\chi^2=1.01$ versus 1.08 for the first
epoch). For the first epoch a BPL is significantly better, with an
F-test-value of 19.1 (chance probability $P=1.6\times10^{-8}$, see
also Table~1).

We then fit the first epoch only with a physically motivated SBPL
spectrum described in \citet{GS02}, with a fixed sharpness of the
break\footnote{This value corresponds to the cooling-break for our
inferred photon index and external density profile \citep{GS02}.},
$s=0.85$, and including the broadband LAT UL. We performed two fits:
(i) keeping the two power-law indices free, and (ii) requiring them to
differ by $\Delta\Gamma=0.5$ according to the synchrotron radiation
theoretical expectation \citep{GS02}. The SBPL fit was better
(Table~1) and is shown at the top panel of Fig.$\;$\ref{fig:sed}, together
with the LAT ULs, as well as the extrapolation of the LAT lightcurve
to this epoch; the extrapolation was not used in the fit but plotted
for comparison with the model. Both are consistent with the SBPL fit
-- the curvature in the \nustar\ data is also clearly exhibited in the
inset in the top panel. The lower panel shows the SED with the second
\nustar\ epoch fit with a PL and with the first epoch fit shifted and
superposed on the plot; although the data do not constrain such a fit,
they are consistent with it. Finally, we performed broadband fits
removing the \nustar\ data (including only optical, \swift/XRT,
\swift/UVOT data, and \fermi/LAT ULs) and found that the break
energies could not be constrained. Therefore, the \nustar\ data are
{\it essential in constraining the shape of the broadband spectra}.

Our results are broadly consistent with those of \citet{Perley13} who
derived radio to GeV afterglow spectra of GRB$\;$130427A covering
$0.007-60\;$days after trigger. Their results also suggest that the
forward shock emission indeed dominates at or above the optical during
our \nustar\ epochs.

\begin{deluxetable*}{cccccccccc}
\tabletypesize{\scriptsize}
\tablewidth{0pt} 
\tablecaption{Broadband Spectral Fits during the \nustar\ Epochs} 
\tablehead{
  \colhead{Model\tablenotemark{1}} & \colhead{Epoch} & \colhead{O+X\;\tablenotemark{2}} & \colhead{N\;\tablenotemark{2}} & \colhead{L\;\tablenotemark{2}} & 
\colhead{$\Delta \Gamma$\;\tablenotemark{2}} & \colhead{$\Gamma_1$} & \colhead{$\Gamma_2$} & \colhead{$E_c$\;\tablenotemark{2}} & \colhead{$\chi^2/$d.o.f.} }
\tableline

\startdata
PL & 1 & yes & yes & - & - & $1.72\pm0.02$ & - & - & 457.6/422\tablenotemark{a}  \\
PL & 2 & yes & yes & - & - & $1.77\pm0.02$ & - & - & 105.1/104\tablenotemark{b}  \\
BPL & 1 & yes & yes & - & free & $1.70\pm0.01$ & $1.89^{+0.08}_{-0.04}$ & $9.3^{+2.3}_{-1.4}$  & 419.3/420\tablenotemark{c}  \\
BPL & 2 & yes & yes & - & free & $1.77$ & - & - & -\tablenotemark{d}  \\
BPL & 1 & yes & yes & - & 0.5 & $1.71\pm0.01$ & 2.21 & $17\pm1$ & 428.5/421\tablenotemark{e}  \\
BPL & 2 & yes & yes & - & 0.5 & $1.77\pm0.01$ & 2.27 & $32^{+14}_{-8}$  & 103.7/103\tablenotemark{f} \\
\tableline
\multicolumn{10}{c}{Fits to Optical+X-ray+\nustar\ +LAT confirm presence of break and demonstrate best fit physical model} \\
\tableline
PL & 1 & yes & yes & UL\tablenotemark{3} & - & $1.72\pm0.01$ & - & - & 489.1/434\tablenotemark{g}  \\
{\bf PL}\tablenotemark{4} & {\bf 2} & {\bf yes} & {\bf yes} & {\bf UL}\tablenotemark{3} & {\bf -} & {$\bf 1.76\pm0.01$} & {\bf -} & {\bf -} & {\bf 130.6/116}\tablenotemark{a}  \\
BPL & 1 & yes & yes & UL\tablenotemark{3} & free & $1.70\pm0.01$ & $1.91\pm0.03$ & $9.4^{+1.4}_{-0.9}$ & 428.5/432\tablenotemark{h}  \\
SBPL & 1 & yes & yes & UL\tablenotemark{3} & free & $1.69\pm0.01$ & $2.91^{+0.53}_{-0.49}$ & $96^{+51}_{-25}$ & 422.7/430\tablenotemark{i}  \\
{\bf SBPL}\tablenotemark{4} & {\bf 1} & {\bf yes} & {\bf yes} & {\bf UL}\tablenotemark{3} & {\bf 0.5} & {$\bf 1.67\pm0.01$} & {\bf 2.17} & {\bf $70^{+59}_{-31}$} & {\bf 427.7/429}\tablenotemark{j}
\enddata

\tablenotetext{1}{PL=Power Law, BPL=Broken Power Law, SBPL=Smoothly-Broken Power Law}
\tablenotetext{2}{O+X$\;$=$\;$Optical+{\it Swift}/XRT$\;$+$\;${\it Swift}/UVOT; N$\;$=$\;${\it NuSTAR}; L$\;$=$\;${\it Fermi}/LAT; $\Delta\Gamma=\Gamma_2-\Gamma_1$; $E_c=\;$break energy in keV.}
\tablenotetext{3}{This fit includes the LAT spectra}
\tablenotetext{4}{This spectral fit is shown in Fig.$\;$\ref{fig:sed}}
\tablenotetext{a}{PL is an adequate fit}
\tablenotetext{b}{PL is an good fit}
\tablenotetext{c}{BPL is a better fit than PL, F-test=19.1 ($P=1.6\times10^{-8}$)}
\tablenotetext{d}{Cannot constrain break}
\tablenotetext{e}{BPL ($\Delta\Gamma=0.5$) is a better fit than PL, F-test=28.5 ($P=1.5\times10^{-7}$)}
\tablenotetext{f}{BPL ($\Delta\Gamma=0.5$) is not significantly better fit than PL, F-test=1.3 ($P=0.25$)}
\tablenotetext{g}{PL is not a very good fit}
\tablenotetext{h}{BPL is a better fit than PL, F-test=30.5 ($P=3.9\times10^{-13}$), break is needed}
\tablenotetext{i}{SBPL is a better fit than PL, F-test=16.9 ($P=7.2\times10^{-13}$)}
\tablenotetext{j}{\bf{SBPL is a better fit than PL, F-test=12.3 ($P=3.5\times10^{-11}$)}}

\label{table:popdesc}

\end{deluxetable*}

\section{Discussion}
\label{sec:disc}

We have shown above that the \nustar\ data are consistent with a power
law in time and frequency below the cooling-break photon energy
$E_c$, $F_\nu\propto t^{-\alpha}\nu^{-\beta}$ with $\alpha=1.30\pm
0.05$ and $\beta\equiv\Gamma_1-1=0.69\pm0.01$ (see Table~1). For
the likely power-law segment (G from \citealt{GS02}) of the
synchrotron spectrum this implies a power-law index of the external
medium density, $\rho_{\rm ext}\propto{}R^{-k}$, where $R$ is the
distance from the central source, of
$k=4/[1+1/(2\alpha-3\beta)]=1.4\pm0.2$. Correspondingly, the
cooling-break energy scales as
$E_c\propto{}t^{(3k-4)/(8-2k)}=t^{0.05\pm0.12}$, {\it i.e.}, it is
expected to remain constant (which is consistent with our spectral
fits, the difference between the two epochs being less than
2$\sigma$). The value we obtain for $k$ is intermediate between a
uniform interstellar medium ($k=0$) and a canonical massive-star wind
($k=2$), possibly indicating that the massive GRB progenitor has
produced an eruption ({\it e.g.}, is opacity driven) prior to its
core-collapse, which alters the circumstellar density profile
\citep{Fryer06}. Such an eruption might also account for a variable
external density profile, where a transition from a flatter profile to
a steeper one might be responsible for the steepening of the
optical-to-X-ray lightcurves after several hours
\citep{Ackermann13,Laskar13}. The density profile might have been
relatively steep ($k\sim1-2$) during the first few hundred seconds,
shortly after the outflow deceleration time, possibly accounting for
the early reverse shock emission \citep{Laskar13,Perley13}.

The \nustar\ power-law distributions in time and frequency support an
afterglow synchrotron origin \citep{KBD09,KBD10,Ghisellini10}. 
Synchrotron radiation models predict a maximum synchrotron photon
energy, $E_{\rm{}syn,max}$, derived by equating the electron
acceleration and synchrotron radiative cooling timescales, assuming a
single acceleration and emission region \citep{GFR83,deJager96,KR10,PN10}. 
In the context of late-time \fermi/LAT high-energy photons, this
was first briefly mentioned as a problem for a synchrotron origin for
GRB\,090902B \citep{Abdo09}, and later discussed more generally and in
depth by \citet{PN10}. The long-lasting ($\sim$1$\;$day)
\fermi/LAT afterglow included a 32$\;$GeV photon after 34$\;$ks, and
altogether five $>$30$\;$GeV photons after $>$200$\;$s. All five
significantly exceed $E_{\rm{}syn,max}$, by factors of 6--25 for
$k=0$, and 9--20 for $k=2$ \citep[using Eq.$\;$(4) of][]{PN10}.  This
led to suggestions that the \fermi/LAT high-energy photons were not
synchrotron radiation, but instead arose from a distinct high-energy
spectral component
\citep{Ackermann13,Fan13}.

Such a component may arise for example, from synchrotron self-Compton
\citep{Fan13}.  This mechanism was predicted to dominate at high
photon energies at late times \citep{PK00,SE01}, but has rarely been
detected in the late X-ray afterglow \citep{Harrison01,Yost02}. Other
possible origins of the high-energy emission involve long-lived
activity of the central source, producing a late relativistic outflow
that provides seed synchrotron photons or relativistic electrons that
might scatter either their own synchrotron emission or that of the
afterglow shock \citep{FP08}. In GRB$\;$130427A, however, there are no
signs of prolonged central source activity (such as X-ray flares)
beyond hundreds of seconds. Another option is a ``pair echo''
involving TeV photons emitted promptly by the GRB, which pair-produce
with photons of the extragalactic background light; for low enough
intergalactic magnetic fields the resulting pairs can produce
detectable longer-lived GeV emission by up-scattering cosmic microwave
background photons \citep{Plaga95,Takahashi08}. However, in this case
the flux decay rate is expected to gradually steepen and the photon
index to soften, in contrast with observations. A different
possibility is pair cascades, induced by shock-accelerated
ultra-high-energy cosmic rays \citep{DA06}.

For any of these alternative models to work, there needs to be a
transition from synchrotron emission (at low photon energies) to the
alternative model (at high energies). We expect that if a distinct
spectral component dominated the emission at GeV energies, it would
naturally show up in a broad-band SED. By combining optical, XRT,
\nustar\ and \fermi/LAT UL data, we have shown that the SED at
$\sim$1.5$\;$days is perfectly consistent with the theoretically
expected SBPL spectral shape from optical to GeV energies, without any
unaccounted-for flux, and that the flux at all these energies decays
at a similar rate. This strongly suggests a single underlying spectral
component over a wide energy range. For low energies, the most viable
emission mechanism for such a spectral component is synchrotron
radiation, suggesting that the entire SED is produced by synchrotron
emission.

Therefore, our results strongly suggest that the late-time
\fermi/LAT high-energy photons in GRB$\;$130427A are indeed afterglow 
synchrotron radiation, and provide the strongest direct observational
support to date for such an afterglow synchrotron origin of late-time
$>$10$\;$GeV \fermi/LAT photons. As was already pointed out
\citep[e.g.,][]{PN10}, such an origin challenges particle acceleration 
models in afterglow shocks.  In particular, at least one of the
assumptions in estimating $E_{\rm syn,max}$ must be incorrect,
requiring a modification of our understanding of afterglow shock
physics. While many authors were aware of this potential problem, the
NuSTAR results make it much harder to circumvent. One possible
solution may lie in changing the assumption of a uniform magnetic
field into a lower magnetic field acceleration region and a higher
magnetic field synchrotron radiation region
\citep{Kumar12,Lyutikov10}. These might arise for diffusive shock
acceleration (Fermi Type~I) if the tangled shock-amplified magnetic
field decays on a short length scale behind the shock front (where
most of the high-energy radiation is emitted), while the highest
energy electrons are accelerated in the lower magnetic field further
downstream \citep{Kumar12}.

Another possibility is direct linear acceleration in the electric
field of magnetic reconnection layers, which have a low magnetic field
\citep{UCB11,CUB12,CWUB13}. This would require, however, a significant
fraction of the total energy in the flow to reside in magnetic fields
of alternating sign. This is not expected in GRB afterglows, but it
could occur in the magnetic-reconnection induced decay of the tangled
shock-amplified field mentioned above, which initially reaches
near-equipartition values just behind the shock. While the exact
solution is still unclear, our results provide an important challenge
for our understanding of particle acceleration and magnetic field
amplification in relativistic shocks.

\acknowledgements
{\small This work was supported under NASA Contract$\;$NNG08FD60C, and
made use of data from the \nustar\ mission, a project led by CalTech,
managed by JPL, and funded by NASA. We thank the \nustar\ Operations,
Software and Calibration teams for support with the execution and
analysis of these observations. This research has made use of the
\nustar\ Data Analysis Software (NuSTARDAS) jointly developed by the
ASI Science Data Center (ASDC, Italy) and CalTech.  This work made use
of data supplied by the UK \swift\ Science Data Centre at the
University of Leicester. The \fermi/LAT Collaboration acknowledges
support 
from NASA and DOE (US), CEA/Irfu and IN2P3/CNRS (France), ASI and INFN
(Italy), MEXT, KEK, and JAXA (Japan), and the K.A. Wallenberg
Foundation, the Swedish Research Council and the National Space Board
in Sweden. Additional support from INAF in Italy and CNES in France
for science analysis during the operations phase is also gratefully
acknowledged. The Liverpool Telescope is operated by Liverpool John
Moores University at the Observatorio del Roque de los Muchachos of
the Instituto de Astrofisica de Canarias. CGM acknowledges support
from the Royal Society.}

\end{document}